\documentclass[preprint,showpacs,preprintnumbers,superscriptaddress,amsmath,amssymb]{revtex4}
 \usepackage{cases}
 \usepackage{CJK}
 \usepackage{txfonts}
 \usepackage{mathrsfs}
 \usepackage{amssymb}
 \usepackage[dvips]{graphicx}
 \usepackage{dcolumn}
 \usepackage{bm}
 \usepackage{subfigure}
 \usepackage{floatflt}
 \usepackage{float}
 \usepackage{rotating}
 \usepackage{multirow}
 \usepackage[bookmarksnumbered,bookmarksopen,colorlinks,citecolor=blue,linkcolor=blue]{hyperref} %
 \usepackage{epsfig,graphics,psfrag,amsmath}
 \usepackage{sidecap}
\def\gxnu{College of Physics and Technology, Guangxi Normal University, Guilin 541004, China}
\def\gxnp{Guangxi Key Laboratory of Nuclear Physics and Technology, Guangxi Normal University, Guilin 541004, China}
\def\thu{Department of Physics, Tsinghua University, Beijing 100084, China}
\def\cic{Collaborative Innovation Center of Quantum Matter, Beijing 100084, China}
\def\esym{$E_{\rm sym}(\rho)$}
\def\amev{MeV/u}

\begin{document}
 \preprint{00-000}

 \title{A new probe to study symmetry energy at low density by deuteron breakup reaction}

 \author{Xiao Liang}
 \affiliation{\gxnu}

 \author{Li Ou}\email{liou@gxnu.edu.cn}
 \affiliation{\gxnu}\affiliation{\gxnp}

 \author{Zhigang Xiao}\email{xiaozg@tsinghua.edu.cn}
 \affiliation{\thu}\affiliation{\cic}

 \date{\today}

 \begin{abstract}
 The reactions of nucleon and polarized deuteron scattered off a heavy target at large impact parameter with intermediate energies have been investigated by using the improved quantum molecular dynamics model. It is found that, due to the difference effect of isovector potential on proton and neutron, there is a significant difference between the angle distribution of elastic scattering protons and neutrons. To overcome the lack of monochromatic neutron beam, the reaction of polarized deuteron peripherally scattered off the heavy target is used to replace the reaction of individual proton and neutron scattered off heavy target to study the isospin effect. It is found that the distributions of elastic scattering angle of proton and neutron originating from the breakup of deuteron are very similar to the results of the individual proton- and neutron-induced reaction. A new probe more effective and more clean, namely the difference between elastic scattering angle of proton and neutron originating from the breakup of polarized deuteron, is promoted to constrain the symmetry energy at subsaturation density.
 \end{abstract}

 \pacs{21.65.Ef, 25.40.-h, 25.45.-z, 24.10.-i}

 \keywords{nucleon-induced reaction, deuteron-induced reaction, symmetry energy}
 \maketitle


\section{Introduction}

 The equation of state (EOS) of isospin asymmetric nuclear matter is still a hot topic nowadays. Especially, the symmetry energy which characterizes isospin dependence of EOS has received considerable attention in recent years, because of its importance not only to nuclear physics but also to many issues in astrophysics\cite{lba08}, such as the properties of rare isotopes \cite{baran05,niko11}, the stability of superheavy nuclei \cite{dong11},the dynamics of rare isotope reactions \cite{bara05,stei05,latt07}, the structures, composition, and cooling of neutron stars \cite{latt00,horo01,todd05,shar09,prak88}, and the mechanism of core-collapse and explosion of supernovae \cite{bonc81,wata09,shen11}, and so on.

 Unfortunately, because of the well-known difficulties of treating accurately quantum many-body problems and our poor knowledge about the spin-isospin dependence of many-body forces, theoretical predictions for the density dependence of the symmetry energy (\esym) of nuclear matter away from the saturation density show large uncertainties \cite{lba08,brow00}. Many efforts have been devoted to probe and constrain the \esym~ by analyses of terrestrial experiments and astrophysical observations, such as neutron skin \cite{warda10,chen10,gaid12,liu11}, nuclear mass \cite{wang13a,fan14,dani09}, nuclear charge radii \cite{wang13b}, the mass-radius relationship \cite{wang13b}, $\alpha$ decay \cite{dong13}, giant dipole resonance and pygmy dipole resonance \cite{carb10,wiel09,piek83},  isospin diffusion \cite{tsang01,wu02,tsang04,chen05,liu07,sun10,rizz08}, isospin drift \cite{zhang15}, double neutron to proton ratio \cite{chen06,fami06,zhang08,tsang09,kuma11,zhang12,zhang14,xie13}, light charged particle flow \cite{rizz04,yong09,kohl10,gior10,cozm11,wang14}, $\pi^-/\pi^+$ ratio \cite{xiao09,feng10,gao13,hong14,xiao14}, $K^+/K^0$ \cite{lopez07,li05,ferini06,feng13},  and gravitational waves from merging neutron star binaries \cite{stei12,taka14}.

 Although a general consensus on the constraints of \esym~at saturation and subsaturation densities \cite{tsang12,latt14} has been obtained, there is a considerable uncertainty. Further constrain of \esym~at subsaturation is not only necessary for itself but also signifcant for the constrain of \esym~at suprasaturation densities. It is known that the so-called hadronic observables sensitive to \esym~at suprasaturation densities in heavy-ion collisions, $\pi^-/\pi^+$ ratio for example, inevitably suffer from effects of the symmetry energy at low densities during the final-state of reaction. So it is quite important to verify the probed density region of probes. However, one only knows these probes are in general sensitive to the high-density or low-density behaviors of the symmetry energy at certain beam energies and impact parameters. Even some established views about the probes face challenge with the deepening of research. For example, some works show that $\pi^-/\pi^+$ ratio, which is being regarded as one of probes sensitive to \esym~at suprasaturation densities, however probes \esym~around saturation density \cite{yong19}.

 The plight of study on symmetry energy is attributed to two reasons. One is the insufficiency of experimental data, the other is that the extraction of the \esym~from heavy ion collisions (HICs) relies unavoidably on the transport model simulations in the most cases. Although people have organized five international collaborations attempting to find out the origin of different predictions for the same experiments by various transport models and trying to reduce the model uncertainty \cite{zhang18,xu16}, the hopes for thoroughly solving the problem is pretty slim for the foreseeable future. It is thus necessary to propose more symmetry-energy-sensitive probes, which are effective and free from transport model limitations.

 So far, the existing symmetry-energy-sensitive probes are mostly based on HICs. Due to the complexity of HICs, considerable discrepancies in the model outputs lead constraints on the \esym~to be still on the qualitative level. While some types of the direct reaction, like the elastic or quasi-elastic scattering as well as the direct projectile breakup, involve less degrees of freedom in the reaction process and may reduce the difficulties in modeling the collision. The probes of these kind of reactions definitely reflect the information of \esym~at subsaturation densities because the system density is almost unchanged in the collision process. By properly selecting the range of the impact parameter, one can limit the probed density into narrower windows.

 As shown in our previous work, due to the isovector potential, there is a significant difference in the scattering angle between  proton and neutron elastically scattered off a heavy target at large impact parameter \cite{ou08}. And the breakup of polarized deuteron induced on heavy ions provides a novel and more quantitative constraint to the symmetry energy below half of the saturation density. The correlation angle of the proton and neutron from a breakup of deuteron can be a good candidate of probe for \esym~at low density \cite{ou15}. As a follow-up work, in this paper, we promote one more symmetry-energy-sensitive probe, namely the difference between elastic scattering angle of proton and neutron originating from the breakup of deuteron, to constrain the \esym~at low densities.

 The paper is organized as follows. In Sec. II, we briefly introduce the model. In Sec. III, we present the isospin effect in nucleon-induced reactions and polarized deuteron breakup reactions. Finally a brief summary is given in Sec. IV.

\section{Model}

 The improved quantum molecular dynamics (ImQMD) model is an extended version of quantum molecular dynamics (QMD) model
 for the simulations of the heavy ion collisions at intermediate beam energies\cite{aichelin91,wang02,zhang05,ou08}.
 The QMD model has been successfully applied in the study of heavy ion collisions at intermediate energies and also has been applied in the proton-induced collisions and provides consistent description to the experimental data if available \cite{niita95,chiba96,chiba962,ouli07,ouli09}.

 In the ImQMD model, each nucleon is described by a Gaussian wave packet,
 \begin{eqnarray} \label{wp}
  \psi_{i}(\bm{r})=\frac{1}{(2\pi\sigma_{r}^{2})^{3/4}}
  \exp\left[-\frac{(\bm{r}-\bm{r}_{i})^{2}}{4\sigma_{r}^{2}}
  +\frac{i}{\hbar}\bm{r}\cdot\bm{p}_{i}\right],
 \end{eqnarray}
 here $\bm{r}_i$ and $\bm{p}_i$ are the center of the $i$th wave packet in the coordinate and momentum space, respectively, and $\sigma_r$ is the width of wave packet, which satisfy $\sigma_r \cdot \sigma_p = \textstyle{\hbar \over 2}$. By making the Wigner transform on the wave function, the one-body phase space distribution function can be obtained, which read as,
  \begin{eqnarray} \label{Wigner}
  f(\bm{r},\bm{p})=\sum_{i=1}^{A}\frac{1}{(\pi\hbar)^3}
        \exp\left[-\frac{(\bm{r}-\bm{r}_i)^2}{2\sigma_{r}^{2}}\right]
        \times \exp\left[-\frac{(\bm{p}-\bm{p}_i)^2}{2\sigma_{p}^{2}}\right].
 \end{eqnarray}

\begin{figure}[h]
 \centering
 \includegraphics[angle=0,width=0.4\textwidth]{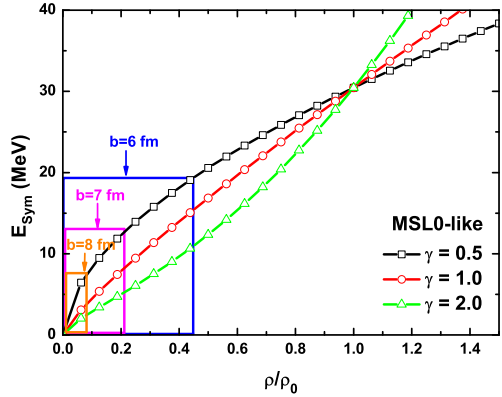}
 \caption{(Color online) The density dependence of symmetry energy given by MSL0-like Skyrme interactions with $\gamma=$0.5, 1.0 and 2.0 adopted.} \label{MSL0}
 \end{figure}

  The time evolution of $\bm{r}_i$ and $\bm{p}_i$ for each nucleon is determined by solving Hamiltonian equations of motion
 \begin{eqnarray}\label{Hequations}
 \dot{\bm{r}}_{i}=\frac{\partial H}{\partial\bm{p}_{i}}, \;\; \;
 \dot{\bm{p}}_{i}=-\frac{\partial H}{\partial\bm{r}_{i}},
 \end{eqnarray}
 where
 \begin{eqnarray}\label{Hamiton}
 H=T+U_{\rm{Coul}}+U_{\rm{loc}},
 \end{eqnarray}
 here, the kinetic energy $T=\sum\limits_{i}\frac{\bm{p}_{i}^{2}}{2m}$, $U_{\rm{Coul}}$ is the Coulomb energy, and the nuclear local potential energy $U_{\rm{loc}} =\int V_{\rm{loc}}[\rho(\bm{r})]d \bm{r}$, where $V_{\rm{loc}}$ is the full Skyrme type potential energy density functional with just the spin-orbit term omitted, which reads
 \begin{align}\label{Vloc}
 V_{\rm loc}=&\frac{\alpha}{2}\frac{\rho ^{2}}{\rho _{0}}+\frac{\beta }{\eta +1}%
 \frac{\rho ^{\eta +1}}{\rho _{0}^{\eta }}
 +\frac{g_{\rm sur}}{2\rho _{0}}\left(\nabla \rho \right)^{2}
 +\frac{g_{\rm sur,iso}}{\rho_{0}}[\nabla(\rho_{n}-\rho_{p})]^{2}\nonumber\\
 &+g_{\rho\tau}\frac{\rho^{8/3}}{\rho_{0}^{5/3}}+(A\rho+B\rho^{\gamma}+C\rho^{5/3})\delta^{2}\rho,
 \end{align}
 where $\rho$, $\rho_n$ and $\rho_p$ are the saturation density, neutron and proton densities, respectively, and the isospin asymmetry $\delta = (\rho_n -\rho_p)/(\rho_n +\rho_p)$. All parameters in Eq. \eqref{Vloc} can be derived from the standard Skyrme interaction parameters \cite{ou08}. To mimic the strong variation of \esym~as well as keep the isoscale part of EOS unchanged, the volume symmetry potential energy term (corresponding to the last term in Eq. \eqref{Vloc}) is replaced with the form of $\textstyle{C_{\rm s,p}\over 2}\left( \textstyle{\rho \over \rho_0} \right)^{\gamma}\rho$, by setting $A_{\rm sym} = C_{\rm sym} = 0$ and $B_{\rm sym} = \textstyle{C_{\rm s,p} \over 2}$.
 Then the symmetry energy is written as
 \begin{eqnarray}\label{Esym}
 E_{\rm sym}(\rho)=\frac{C_{\rm s,k}}{2}\left( \frac{\rho}{\rho_0} \right)^{2/3}
 + \frac{C_{\rm s,p}}{2}\left( \frac{\rho}{\rho_0} \right)^{\gamma},
 \end{eqnarray}
 where $C_{\rm s,k}$ and $C_{\rm s,p}$ are symmetry kinetic and potential energy parameter,
 respectively. 
 The Skyrme parameter set MSL0 \cite{chen10}, one of Skyrme parameter sets which best satisfy the current understanding of
 the physics of nuclear matter over a wide range of applications \cite{dutra12}, is used in this work.
 By using various $\gamma$, one can get MSL0-like Skyrme interactions with various \esym. In figure \ref{MSL0}, the density dependence of symmetry energy with MSL0-like Skyrme interaction and $\gamma=$0.5, 1.0 and 2.0 adopted are presented. The boxes indicate the probed density windows of observable in this work, which is discussed in the following text.

 While the initialization of the heavy-ion is done as usual as that in \cite{wang02}, the deuteron is semiclassically initialized in a simplified scheme as following. The neutron-to-proton direction is taken as the long symmetric axis (LSA). The initial distance between neutron and proton is set to $3\pm \Delta r$ fm, where $\Delta r$ is a random value in range of 0-0.25 fm. The spatial and momentum coordinates perpendicular to LSA are set to zero. The direction of the momentum is initially set to be opposed for neutron and proton along LSA, and the initial magnitude of the momentum are sampled randomly to obtain a stable deuteron until 100 fm/$c$, namely the root mean square radius of deuteron keeps $2.1 \pm 0.5$ fm, where 2.1~fm is the experimental value for the root mean square radius of deuteron \cite{sick96}. By rotating the LSA randomly or onto a certain direction, one can mimic the unpolarized or pre-oriented deuteron beam as initial state, respectively. For this simplification, the initial distance between the mass centers of projectile and the target is set to 25 fm, then the deuteron will soon enter the target potential field in 30 fm/c for reactions with beam energy of 100 \amev.

\section{Results and discussions}
 In this section, we illustrate and discuss the dynamical isospin effects in nucleon-induced reactions and deuteron-induced reactions.

\subsection{Isospin effects in nucleon-induced reactions}
\begin{figure}[h]
 \centering
 \includegraphics[angle=0,width=0.4\textwidth]{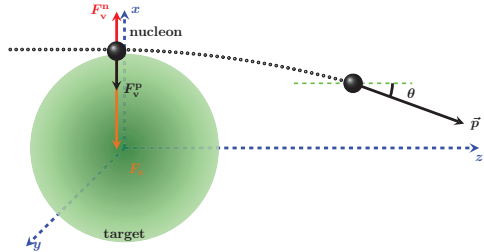}
 \caption{(Color online) The schematic view of a nucleon peripherally scattered off a heavy target.} \label{NIR}
 \end{figure}

\begin{figure}[h]
 \centering
 \includegraphics[angle=0,width=0.4\textwidth]{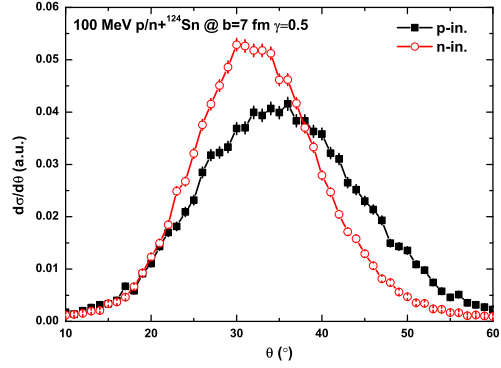}
 \caption{(Color online) The angular distributions of elastic scattering protons and neutrons in nucleon-induced reaction on $^{124}$Sn at $E = 100$ MeV and $b = 7$ fm.} \label{pndis}
 \end{figure}

 When a nucleon peripherally passes by a heavy target nuclei, as shown by the cartoon in figure \ref{NIR}, the nucleon experience nuclear force, and Coulomb force $F_{\rm c}$ for the proton. While the isoscalar nuclear force $F_{\rm s}$ is attractive to both proton and neutron, the isovector force $F_{\rm v}$, is attractive to proton and repulsive to neutron at the subsaturation density environment. This dynamical isospin effect should make opposite effects on the elastic scattering angle of protons and neutrons, and leads to the disparity between the angular distributions of elastic scattering protons and neutrons for the same incident energy and initial geometry.
 This conjecture has already been verified in our previous work, which can be referred to Ref. \cite{ou08} in detail. From the results of the angular distribution of elastic scattering protons and neutrons in the proton or neutron-induced on $^{124}$Sn at $E = 100$ MeV and $b = 7$ fm with the same symmetry energy parameter $\gamma=0.5$ adopted in the calculations, as shown in figure \ref{pndis}, one can see clearly that the elastic scattering protons trend to emit into larger angle while the elastic scattering neutrons trend to emit into smaller angle.

\begin{figure}[h]
 \centering
 \includegraphics[angle=0,width=0.23\textwidth]{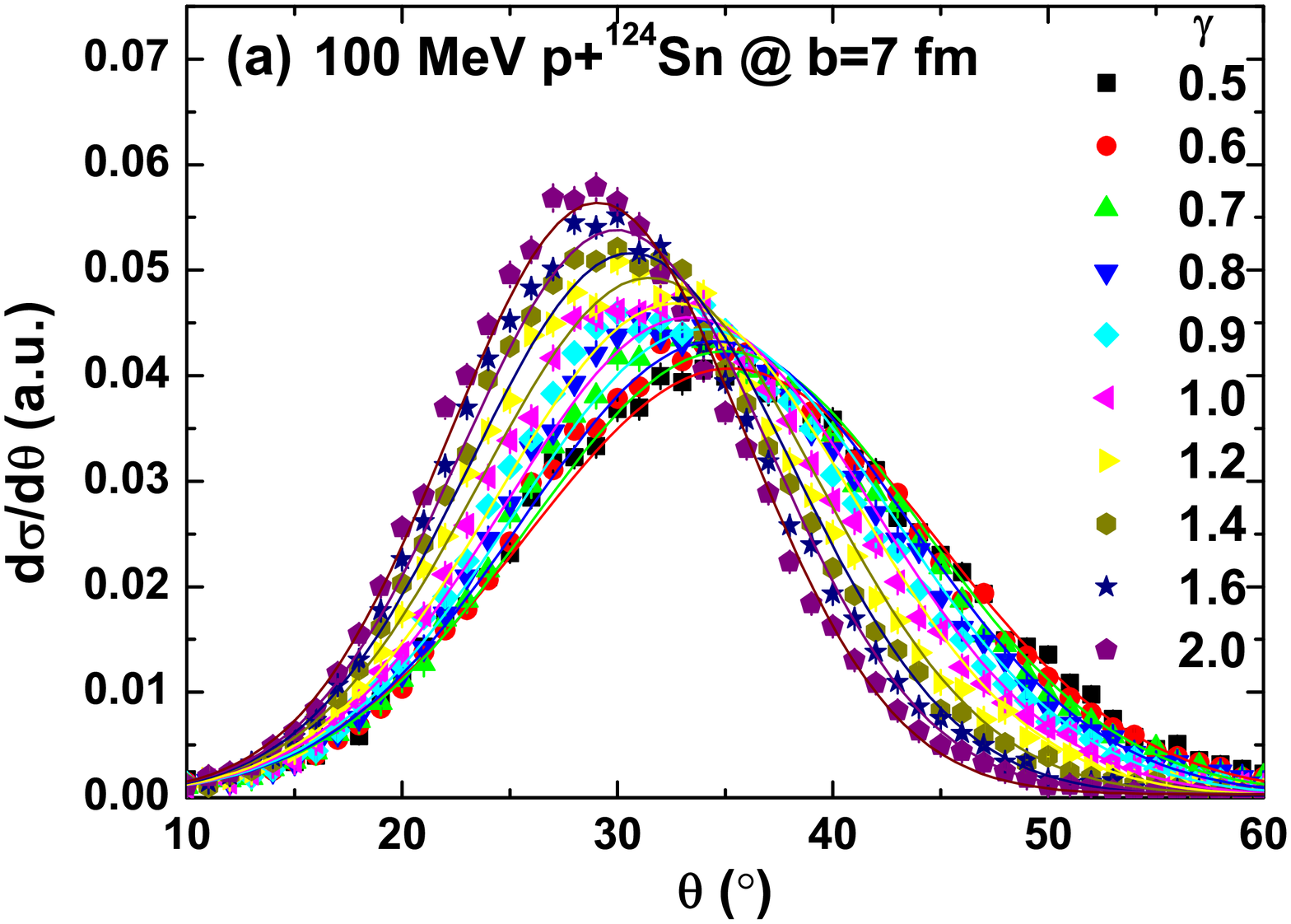}
 \includegraphics[angle=0,width=0.23\textwidth]{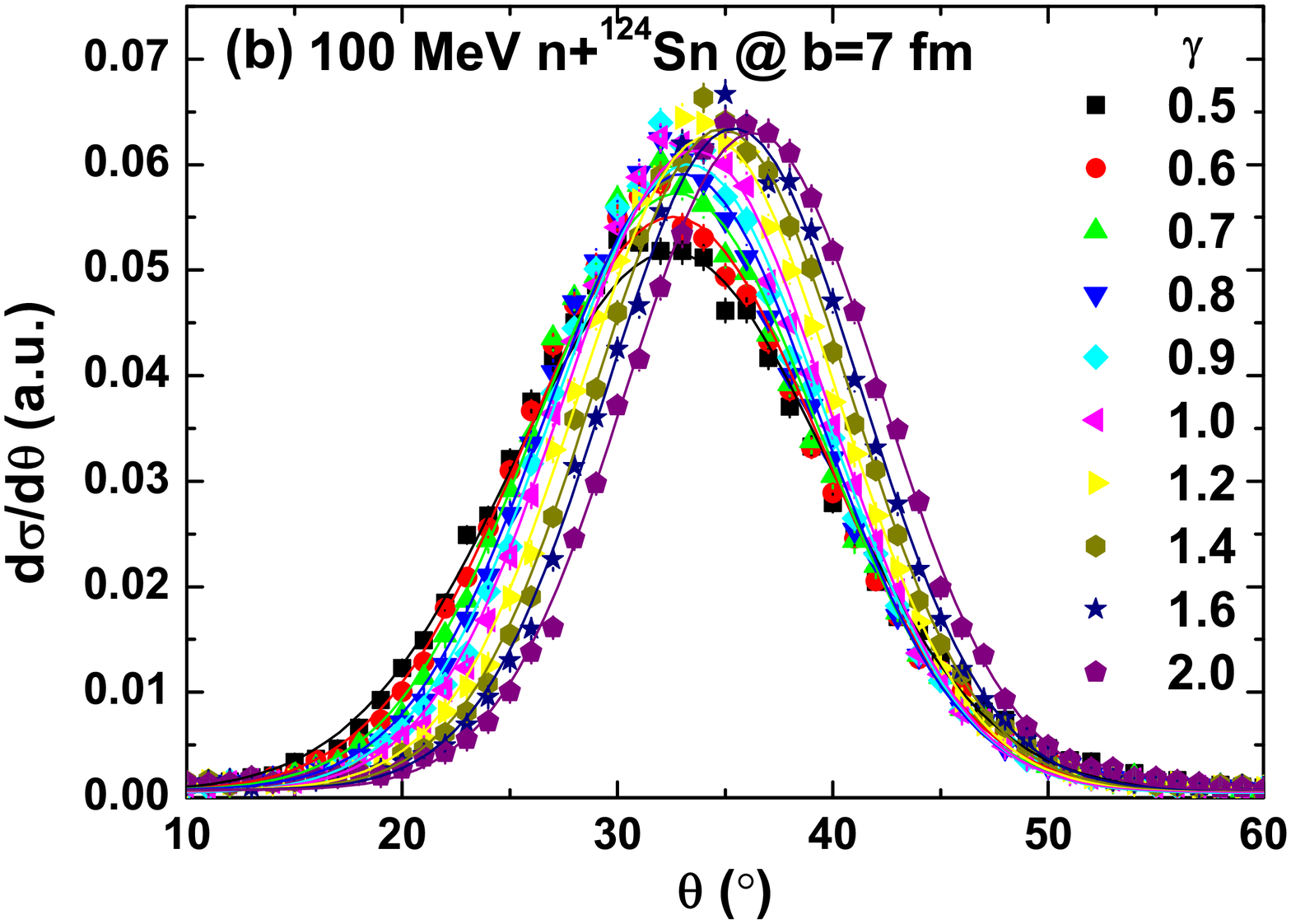}
 \caption{(Color online) The angular distributions of elastic scattering protons (left) and neutrons (right) in corresponding nucleon-induced reaction on $^{124}$Sn at $E = 100$ MeV and $b = 7$ fm, with various \esym~adopted in calculations. The curves are the results by fitting the calculations with Gaussian function.} \label{pngamma}
 \end{figure}

\begin{figure}[h]
 \centering
 \includegraphics[angle=0,width=0.45\textwidth]{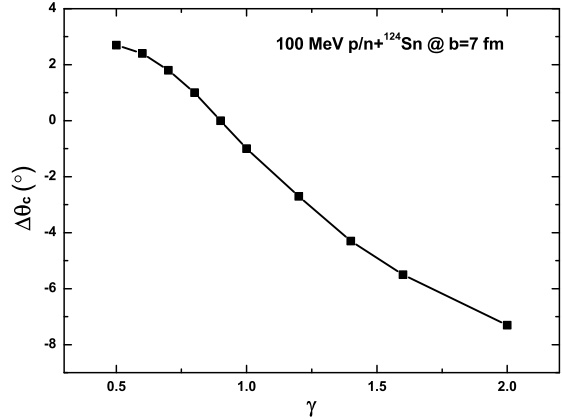}
 \caption{The $\gamma$ dependence of the difference between the location of peak of distribution of elastic scattering protons and neutrons in corresponding nucleon elastically scattered off $^{124}$Sn at $E = 100$ MeV and $b = 7$ fm.} \label{pnxde}
 \end{figure}

 Naturally, the \esym~will affect the elastic scattering angle of protons and neutron. The calculation results of the angular distribution of emitted nucleons with various \esym~($\gamma=0.5-2.0$) adopted are shown in figures \ref{pngamma}.
  The effect of stiffness of symmetry energy on angle distribution is obvious. To quantify the angle distribution in connection with the stiffness of symmetry energy, the angle distributions are fitted with Gaussian distribution function of $\theta$ as
 \begin{equation}\label{Gauss}
 \sigma=\sigma_{0}+\frac{A}{W\sqrt{\pi/2}}e^{-2 \frac{(\theta-\theta_{c})^{2}}{W^{2}}}.
 \end{equation}
 which are presented in figure \ref{pngamma} by corresponding curves.
  \begin{figure}[h]
 \centering
 \includegraphics[angle=0,width=0.45\textwidth]{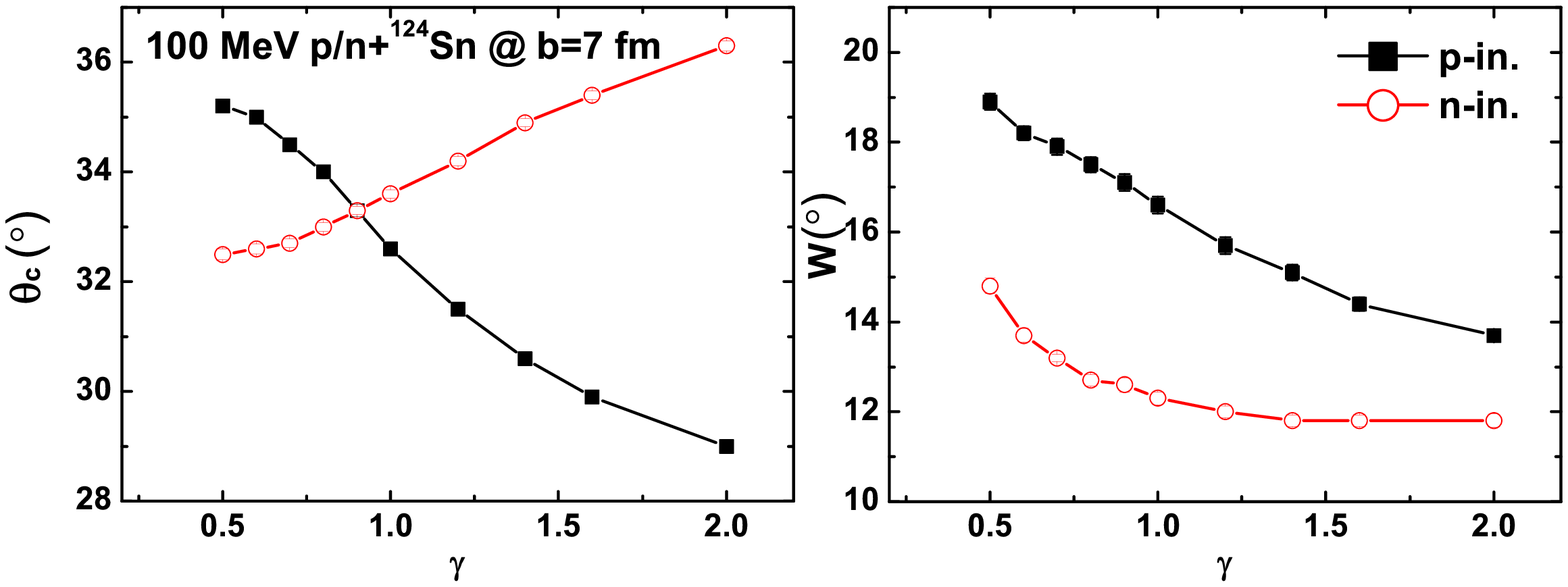}
 \caption{(Color online) The $\gamma$ dependence of the locations of peaks of distribution $\theta_{\rm c}$ (left) and the width of distribution $W$ (right) for protons and neutrons elastically scattered off $^{124}$Sn at $E = 100$ MeV and $b = 7$ fm.} \label{pnxw}
 \end{figure}
 Then the locations of peaks of distribution $\theta_{\rm c}$ and the width of distribution $W$ as a function of $\gamma$, as shown in figure \ref{pnxw}, can be used to study \esym. One can see that, with symmetry energy becomes stiffer, the locations of peaks of distribution of protons $\theta_{\rm c}^{\rm p}$ becomes smaller, which sensitivity to $\gamma$ (from 0.5 to 2.0, similarly hereinafter) is about 25\%; The location of peak of distribution of neutrons $\theta_{\rm c}^{\rm n}$ becomes larger, which sensitivity to $\gamma$ is about 12\%. With symmetry energy becomes stiffer, the width of distribution for both protons and neutrons becomes smaller, which sensitivities are about 35\% and 22\%, respectively. The sensitivities of these observables are close to those of existing observables with sensitivities about 20\%. To get observable with higher sensitivity, the difference between the locations of peaks of distributions of protons and neutrons, namely $\Delta \theta_{\rm c}=\theta_{\rm c}^{\rm p}-\theta_{\rm c}^{\rm n}$, can be constructed. The $\gamma$ dependence of $\Delta \theta_{\rm c}$ is presented in figure \ref{pnxde}. One can see that, with symmetry energy turns stiffer, $\Delta \theta_{\rm c}$ changes from positive to negative, the sensitivity of $\Delta \theta_{\rm c}$ to $\gamma$ is about 200\%, which is more highly sensitive than existing observables.

\subsection{Isospin effects in deuteron-induced reactions}

 Since monochromatic neutron beam with high energy is hardly available, the experiment for neutron-induced reactions remains a difficult task. Thanks to the availability of polarized deuteron beam at hundreds \amev~at various running accelerators around the world \cite{lakin55,moro05,moro09,hata97,oka94}, the deuteron, with one proton and one neutron bound loosely at large average separation distance, provides an alternative opportunity to execute ``proton-neutron-induced'' reactions by deuteron-induced reactions.
\begin{figure}[h]
 \centering
 \includegraphics[angle=0,width=0.45\textwidth]{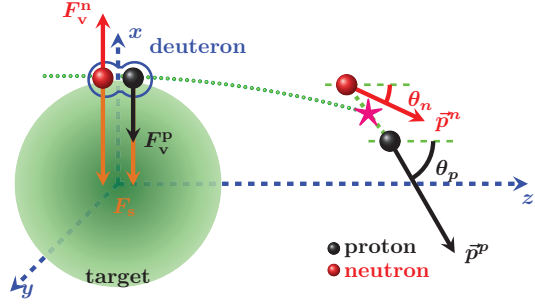}
 \caption{(Color online) The schematic view of a deuteron-induced peripheral collision on a heavy target $^{124}$Sn.} \label{DIR}
 \end{figure}
 If a deuteron breaks without collision when it peripherally passes by the heavy target nuclei, it provides such an mixed proton-neutron beam to probe the isospin effect. As shown by the cartoon in figure \ref{DIR}, the two nucleons in the deuteron experience nuclear force and Coulomb force $F_{\rm c}$, the later of which is repulsive only for the proton. While the isoscalar nuclear force $F_{\rm s}$ is attractive to both nucleons, the isovector force $F_{\rm v}$, is attractive to proton and repulsive to neutron.

 Because of exchange symmetry of the wave function with exchange of n and p, as done in Ref. \cite{ou15}, the simulations is done by mimicking a fully tensor and vector polarized deuteron beam with 50\% possibility for $\vec{r}^{\rm~np} \varparallel  \vec{k}$ and 50\% possibility for $-\vec{r}^{\rm~np} \varparallel \vec{k}$, here $\vec{r}^{\rm~np}$ is the relative vector from neutron to proton and $\vec{k}$ is the particle wave vector. In the following calculations, the LSA of deuteron is preorientated parallel to the beam axis.

\begin{figure}[h]
 \centering
 \includegraphics[angle=0,width=0.23\textwidth]{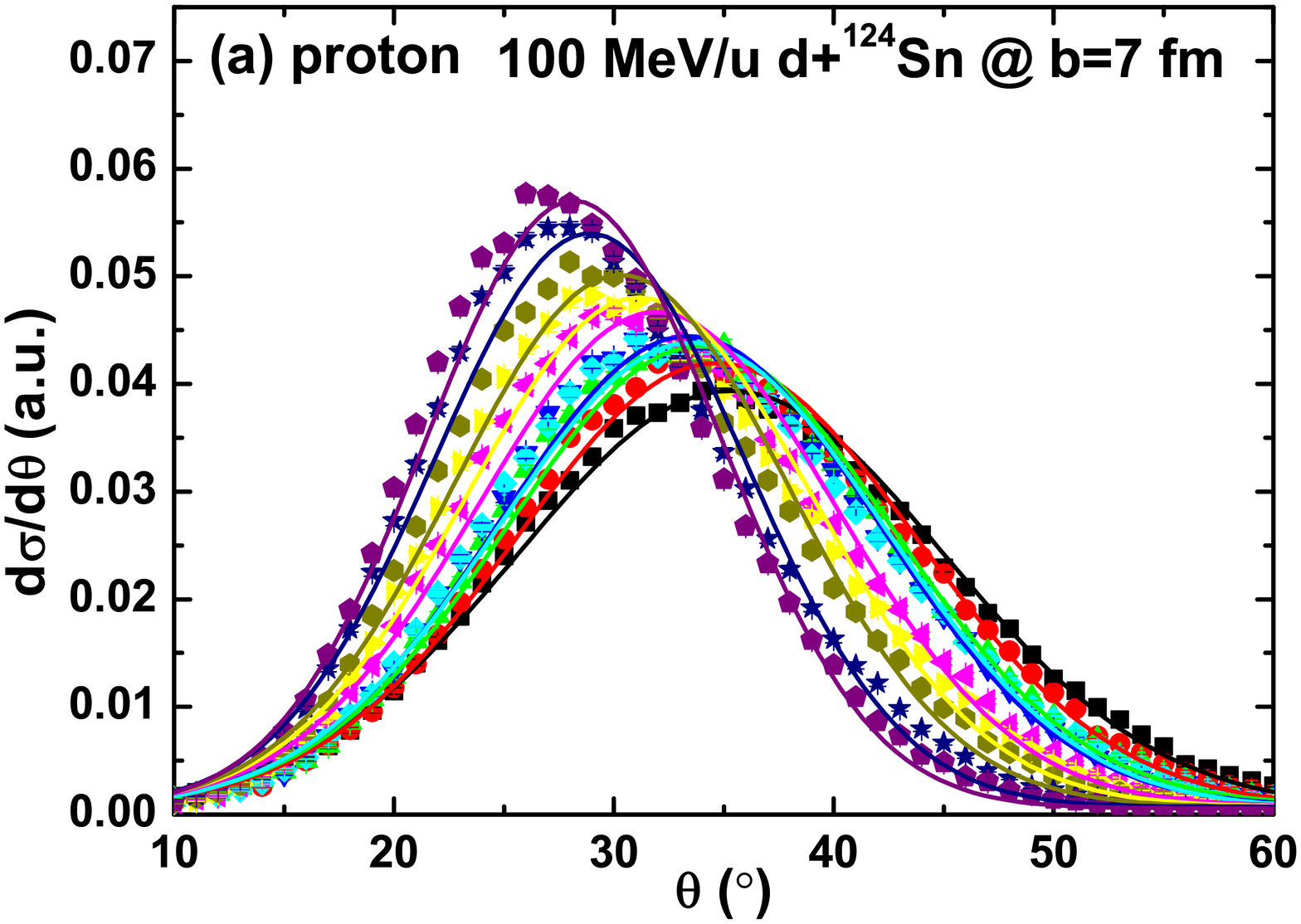}
 \includegraphics[angle=0,width=0.23\textwidth]{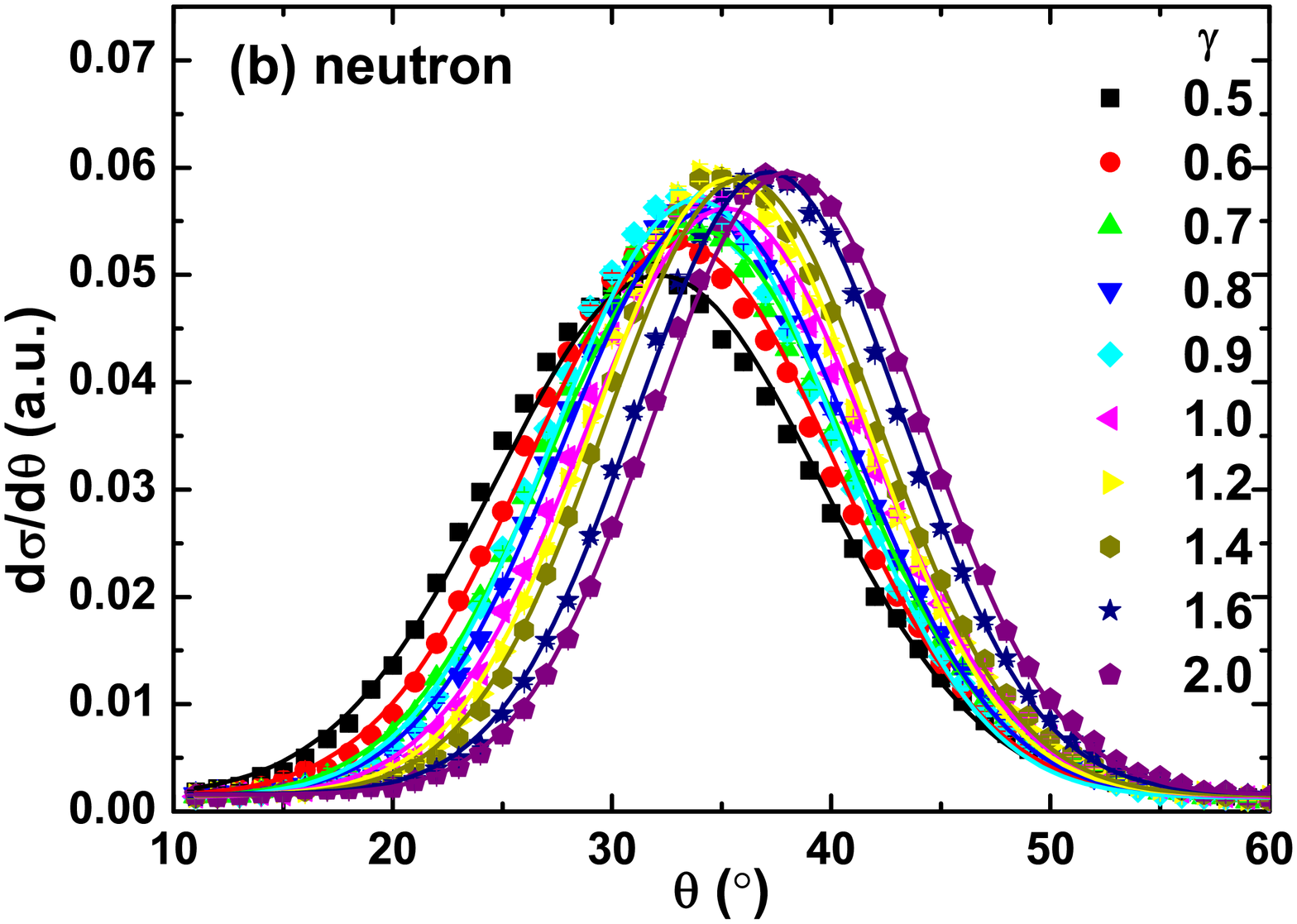}
 \caption{(Color online) The angle distribution of protons (left) and neutrons (right) from breakup of deuterons elastically scattered off $^{124}$Sn with 100 \amev~and $b=7$ fm.} \label{dpn}
 \end{figure}

\begin{figure}[h]
 \centering
 \includegraphics[angle=0,width=0.4\textwidth]{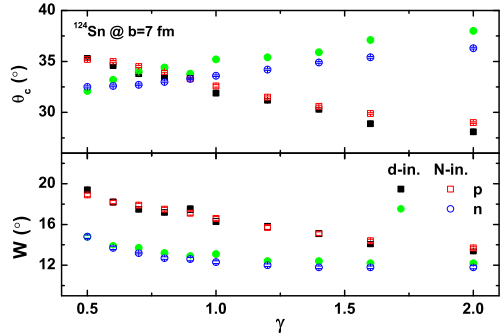}
 \caption{(Color online) The $\gamma$ dependence of the location of peaks $\theta_{\rm c}$ and widths $W$ of distributions of elastic scattering protons and neutrons, where ``d-in.'' denotes $E = 100$ MeV/u deuteron-induced reactions and ``N-in.'' denotes $E = 100$ MeV proton- or neutron-induced reactions on $^{124}$Sn at $b = 7$ fm.} \label{XcWb7}
 \end{figure}

 The angle distribution of protons and neutrons from breakup of polarization deuterons elastically scattered off $^{124}$Sn with 100 \amev~and $b=7$ fm are shown in figures \ref{dpn}. One can see that the behaviors of angle distribution of elastic scattering protons and neutrons originating from the breakup deuterons are quite similar to those in nucleon-induced reactions. Once again, the angle distributions are fitted with Gaussian distribution functions, and the $\gamma$ dependence of the location of peaks $\theta_{\rm c}$ and widths $W$ of distributions are shown in figure \ref{XcWb7}. The locations of peaks $\theta_{\rm c}$ and widths $W$ of distributions for deuteron-induced reactions are so close to those for nucleon-induced reactions. It means that the protons and neutrons from breakup of deuteron in polarized deuteron-induced reactions indeed play ``synchronously'' the corresponding role in nucleon-induced reactions.
%

\begin{figure}[h]
 \centering
 \includegraphics[angle=0,width=0.23\textwidth]{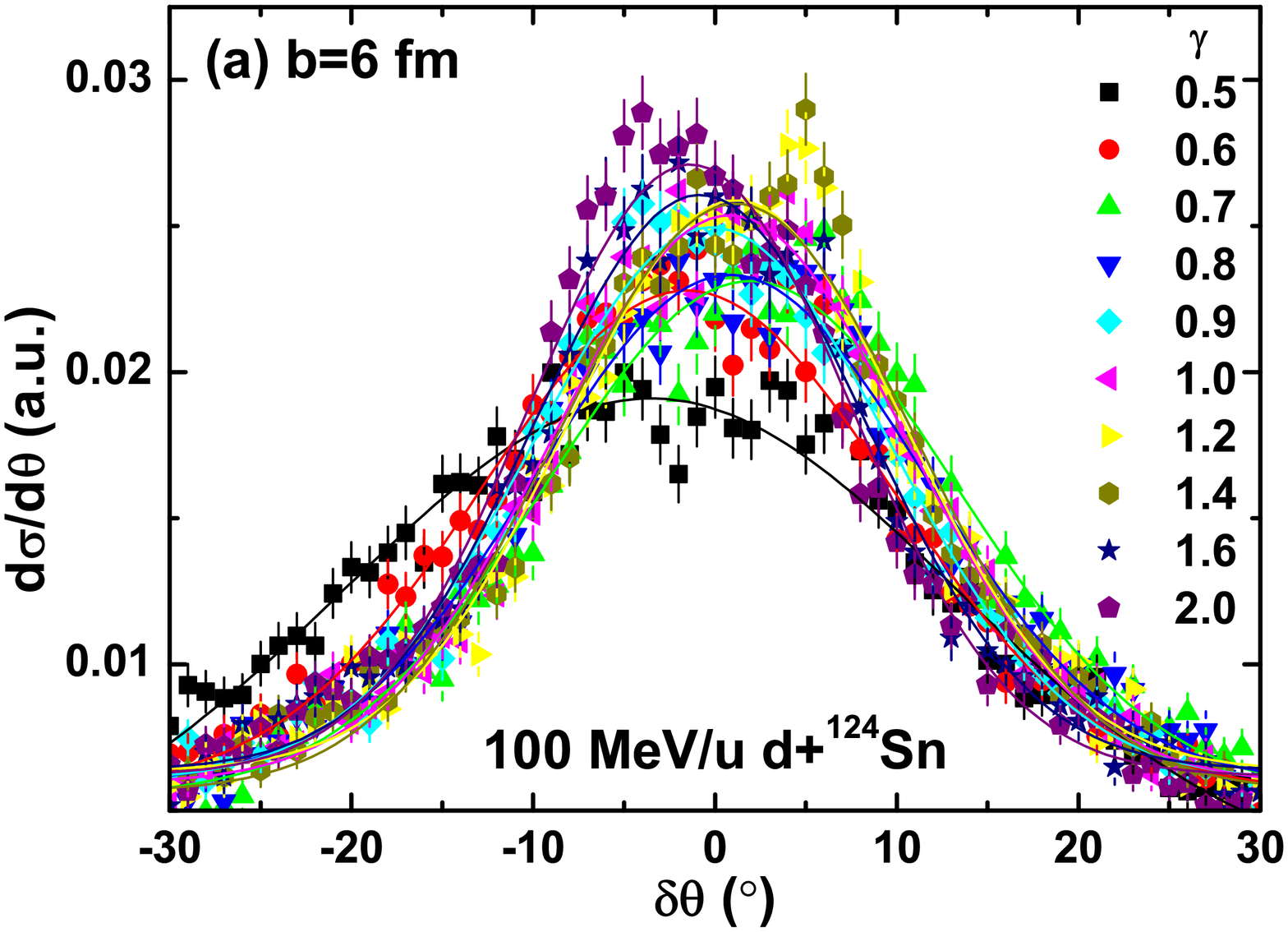}
 \includegraphics[angle=0,width=0.23\textwidth]{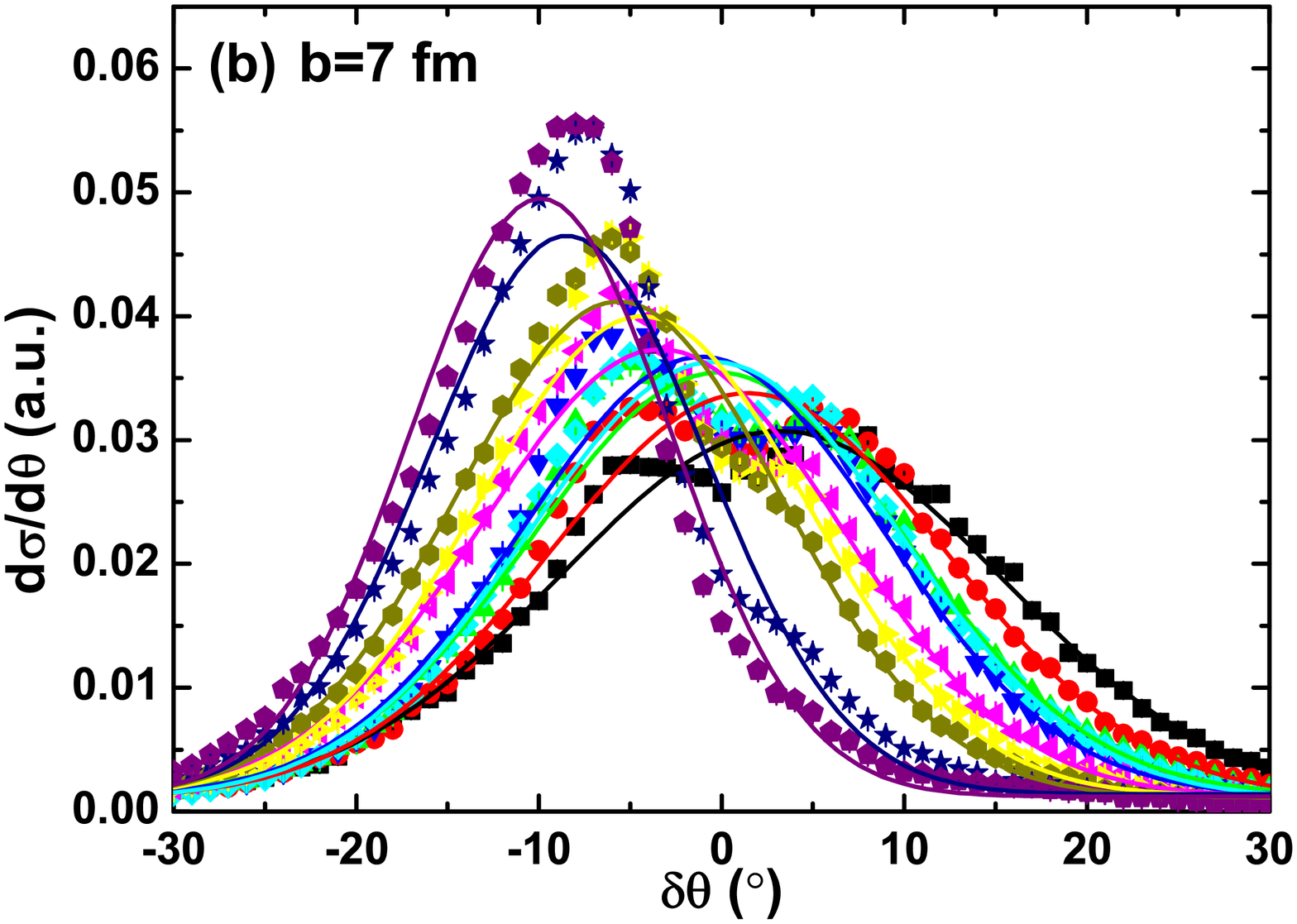}\\
 \includegraphics[angle=0,width=0.23\textwidth]{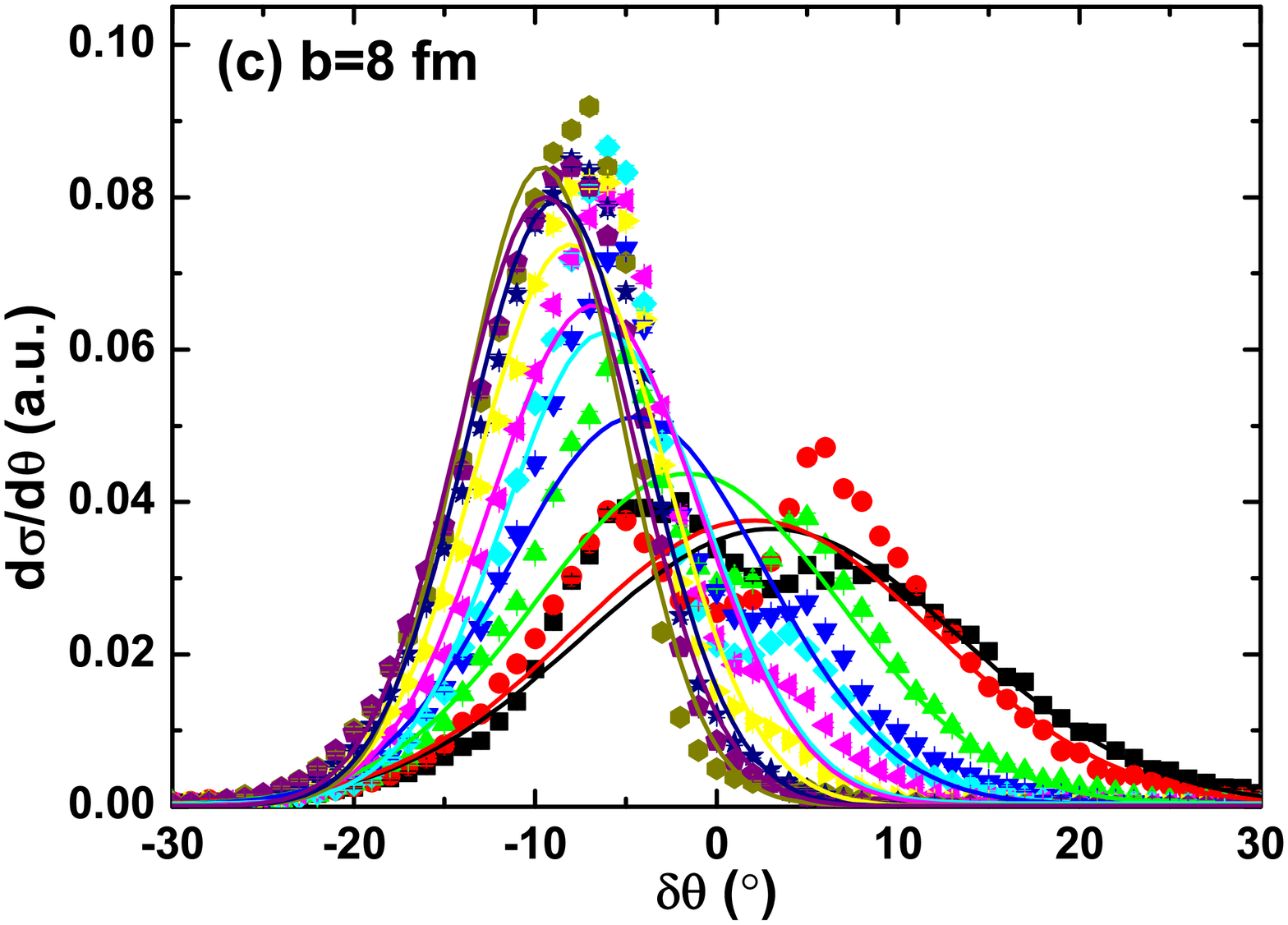}
 \includegraphics[angle=0,width=0.23\textwidth]{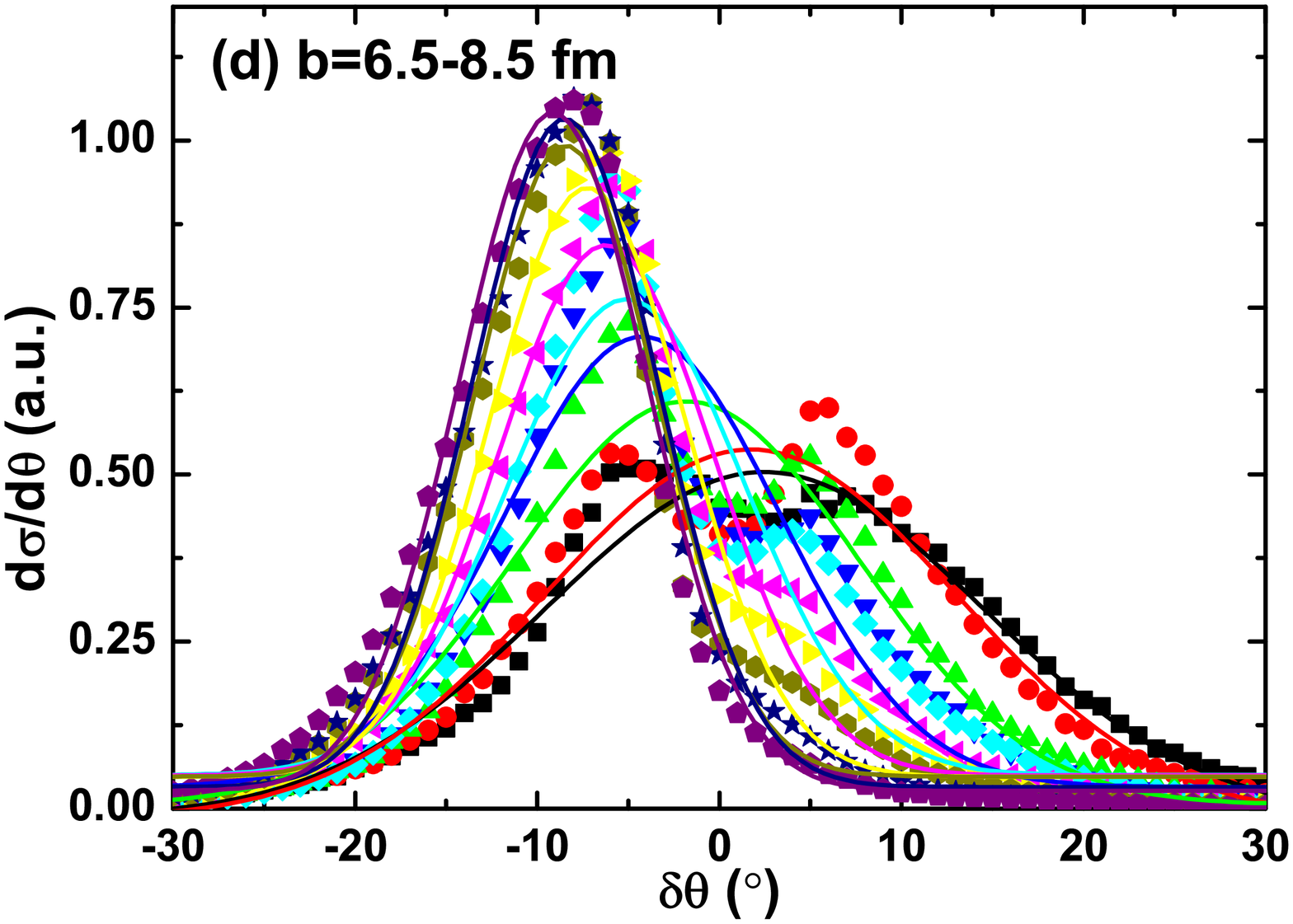}
 \caption{(Color online) The $\gamma$ dependence of the difference between the elastic scattering angle of proton and neutron originating from breakup of deuteron in deuteron-induced reactions on $^{124}$Sn with $E = 100/u$ MeV at $b $=6, 7, 8 and 6.5-8.5 fm.} \label{ddis}
 \end{figure}

 So one can use the difference between the scattering angle of proton and neutron ($\delta\theta= \theta_p-\theta_n$) from breakup of deuteron in each single event, but not the difference between the location of peak of distribution of protons and neutrons, to study the \esym. By this method, the influence from uncertainty of isoscalar potential can be further reduced, because the proton and neutron from deuteron in the same event undergo nearly the same isoscalar potential from target. In figure \ref{ddis}, $\gamma$ dependence of $\delta \theta$ in $E = 100/u$ MeV deuteron-induced reactions on $^{124}$Sn at $b $=6, 7, 8 and 6.5-8.5 fm are presented. For case of $b$=6 fm which deuteron is very close to target, isoscalar potential dominates the scattering, the distinction between the distribution with various \esym~ is not obvious, all centers of $\delta \theta$  distributions locate around zero degree and the widths of distributions are almost the same. With impact parameter increases, the isovector potential effect becomes obvious. The distributions with soft symmetry energy are wider than those with stiff one. The centers of  $\Delta \theta$  distributions with soft symmetry energy trend to locate at positive angles relative to those with stiff one, which trend to locate at negative angles. From the results of $b$=7 and 8 fm, the $\delta \theta$ distribution is not too sensitive to the fine division of impact parameter. That is very helpful to eliminate much of the hardship in experiments and improve the accuracy of constrain on \esym. Finally one can see that the results in peripheral collision mixed with $b=$6.8-8.5 fm still exhibit the sensitivity to \esym.

\begin{figure}[h]
 \centering
 \includegraphics[angle=0,width=0.23\textwidth]{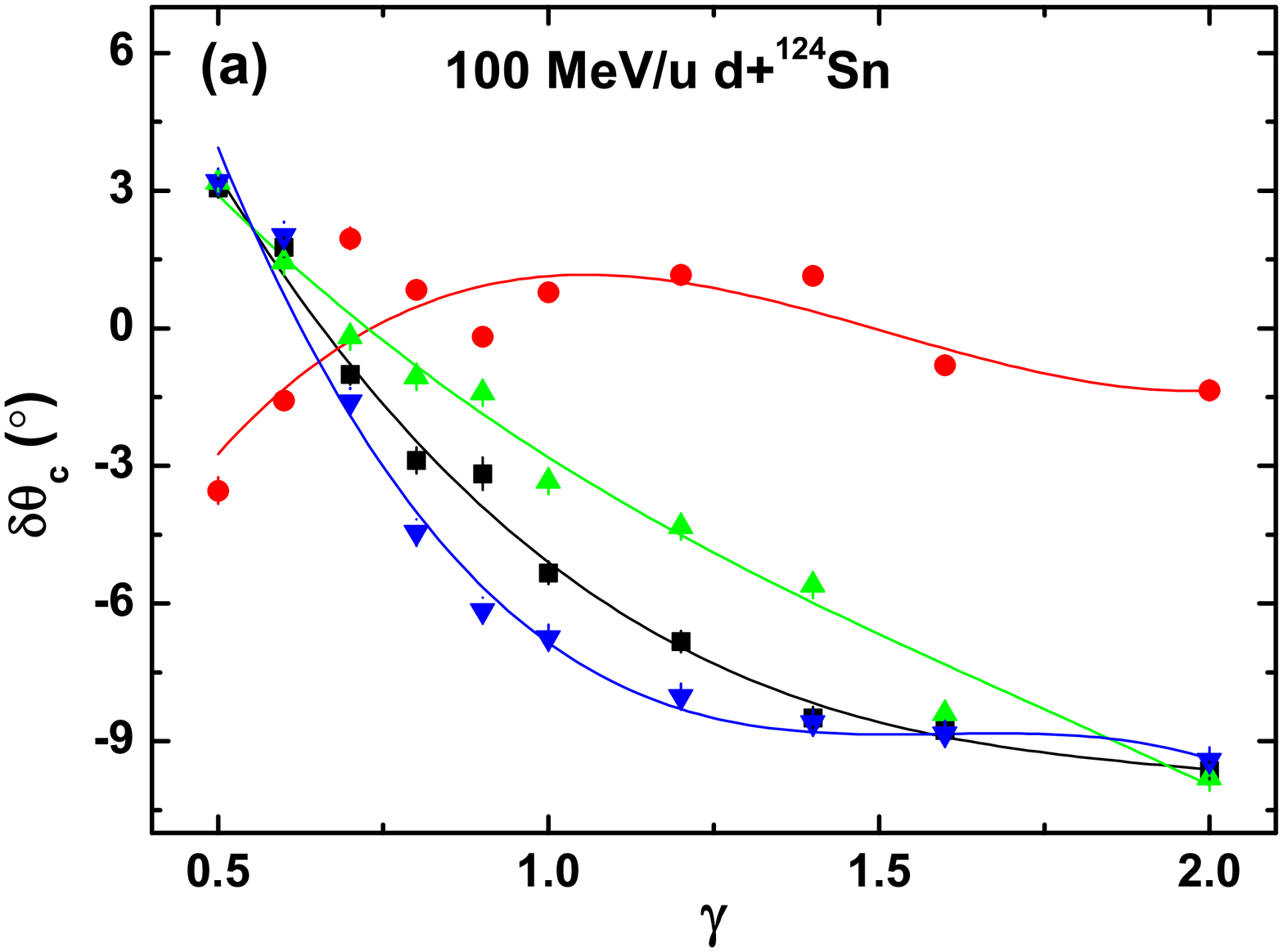}
 \includegraphics[angle=0,width=0.23\textwidth]{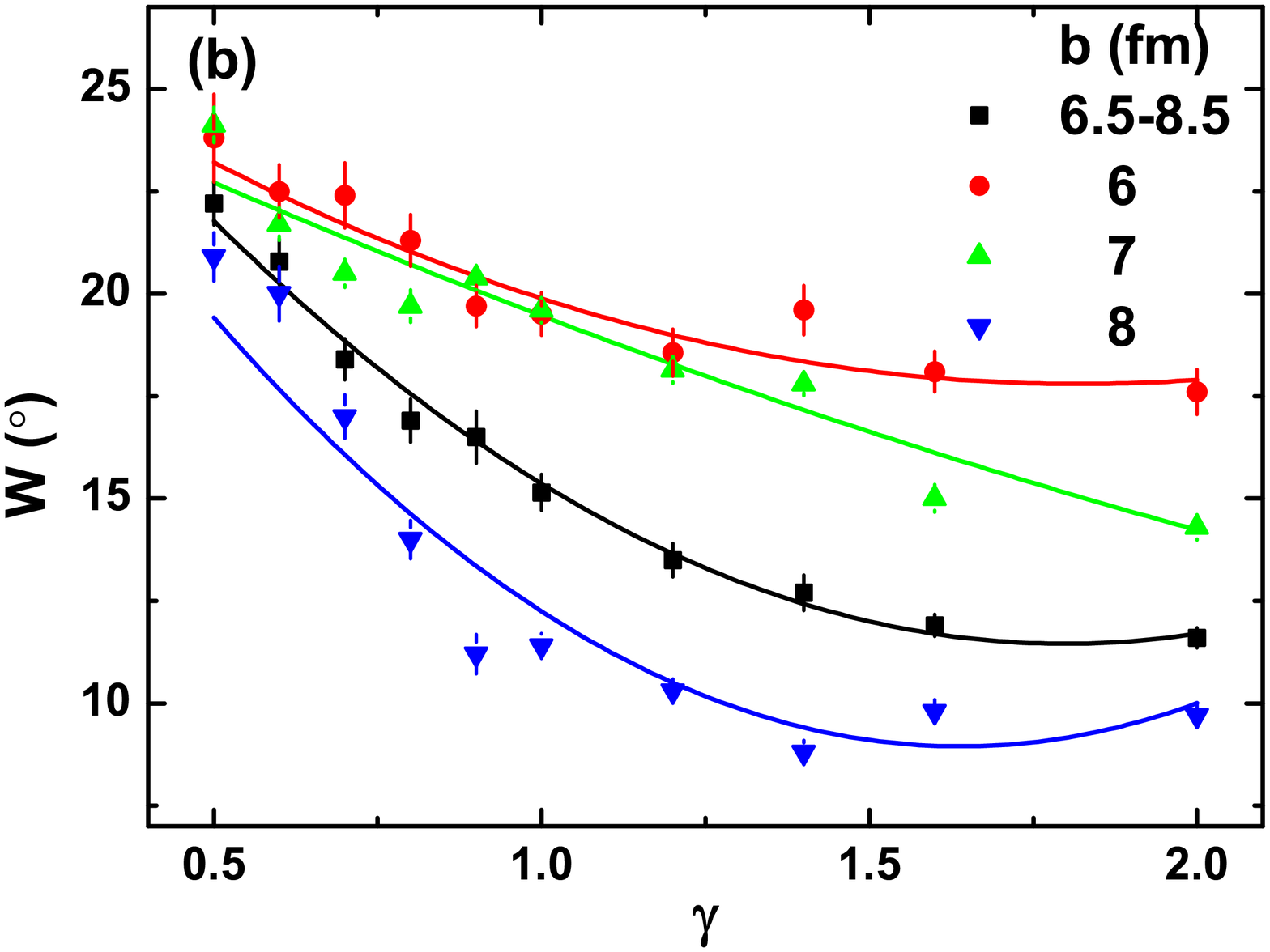}
 \caption{(Color online) The $\gamma$ dependence of the locations of peaks (left) and widths (right) of distributions of difference between the elastic scattering angle of proton and neutron originating from breakup of deuteron in deuteron-induced reactions on $^{124}$Sn with $E = 100$ MeV/u at $b $=6, 7, 8 and 6.5-8.5 fm.} \label{xwall}
 \end{figure}

 Fitting all distributions with Gaussian function, one can get the locations of peaks and widths of $\delta \theta$ distributions. The $\gamma$ dependence of the locations of peaks and widths are presented in figure \ref{xwall}. From the results, one can find that: For small impact parameter, i.e. $b=$6 fm, $\delta \theta_{\rm c}$ is unsensitive to $\gamma$. For large impact parameters, $\delta \theta_{\rm c}$ show strong sensitivity to \esym, decreasing from positive to negative degree with increasing stiffness of symmetry energy. While the widths of distributions $W$ show sensitivity to $\gamma$ for all impact parameters. For $b=$6.5-8.5 fm, the sensitivities of $\delta \theta_{\rm c}$ and $W$ to $\gamma$ are about 400\% and 200\%, respectively. Although $\delta \theta_{\rm c}$ and $W$ are not so sensitive in case of $\gamma>1.2$, it does not impede $\Delta \theta_{\rm c}$ and $W$ from being good candidates to probe \esym, because the very stiff symmetry energy with $\gamma>1.5$ has been ruled out by existing experiments and theories.

\begin{figure}[h]
 \centering
 \includegraphics[angle=0,width=0.4\textwidth]{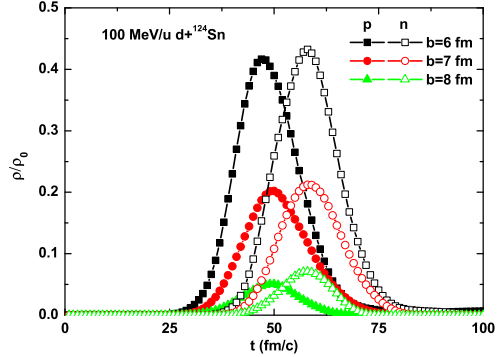}
 \caption{(Color online) The local density experienced by proton and neutron from breakup of deuterons elastically scattered off $^{124}$Sn with 100 \amev~with various impact parameters as function of time.} \label{rho}
 \end{figure}

 Finally, the probed density of this method should be indicated clearly. In the figure \ref{rho}, the local density experienced by protons and neutrons from breakup of deuterons elastically scattered off $^{124}$Sn with 100 \amev~with various impact parameters as function of time are presented. One can see clearly that, the elastically scattering proton and neutron in deuteron peripheral reactions pass through periphery of target nuclei where density below half of saturation density. The probed \esym~windows by this method are shown in the figure \ref{MSL0} by boxes for various impact parameters. It is reasonable to assert that the upper limit of \esym~window is below 0.3$\rho_0$, because the most of collisions with $b=6$ fm (about 94\%) are inelastic scattering due to the collisions with targets.

 \section{Summary}

 Within the ImQMD model, proton-induced and neutron-induced reactions on heavy target with 100 MeV incident energies have been studied. It is found that, due to the isovector potential, there is a significant difference between the elastic scattering angle of proton and neutron in peripheral  reactions. The difference between locations of peaks of distributions of protons or neutrons elastically scattering on heavy targets is very sensitive to density dependence of symmetry energy. To overcome the lack of monochromatic neutron beam, the polarized deuteron peripherally scattered off the heavy target nuclei have been investigated. It is found that the behaviors of angle distribution of elastica scattering protons and neutrons originating from the breakup of deuterons are quite similar to those in corresponding nucleon-induced reactions. So the polarized deuteron scattered off heavy target can be an alternative to the individual proton- and neutron-induced reaction. In terms of the sensitivity and the cleanness, a new probe, namely the difference between elastic scattering angle of proton and neutron originating from the breakup of deuteron, is promoted to be a promising candidate to constraint the symmetry energy at subsaturation density.

 \section{Acknowledgements}
 This work has been supported
 by National Natural Science Foundation of China under Grant Nos.
 11965004, 
 11875174, 11890712, 
 U1867212, 
 11711540016, 
 11847317, 
 and by Natural Science Foundation of Guangxi province under
 Grant No. 2016GXNSFFA380001, 2017GXNSFGA198001,
 Foundation of Guangxi innovative team and distinguished scholar in institutions of higher education,
 and by Tsinghua University Initiative Scientific Research Program.

 


\end{document}